\documentclass[twocolumn]{aastex631}
\usepackage{amsmath}
\usepackage{CJK}
\usepackage{xcolor}
\usepackage{graphicx}

\begin{document}
\begin{CJK*}{UTF8}{gbsn}

\title{New Observations of Solar Wind $1/f$ Turbulence Spectrum from Parker Solar Probe}

\author[0000-0001-9570-5975]{Zesen Huang (黄泽森)}
\affiliation{Department of Earth, Planetary, and Space Sciences, University of California, Los Angeles, CA, USA}
\author[0000-0002-1128-9685]{Nikos Sioulas}
\affiliation{Department of Earth, Planetary, and Space Sciences, University of California, Los Angeles, CA, USA}
\author[0000-0002-2582-7085]{Chen Shi (时辰)}
\affiliation{Department of Earth, Planetary, and Space Sciences, University of California, Los Angeles, CA, USA}
\author[0000-0002-2381-3106]{Marco Velli}
\affiliation{Department of Earth, Planetary, and Space Sciences, University of California, Los Angeles, CA, USA}
\affiliation{International Space Science Institute, Bern 3012, CH}

\author[0000-0002-4625-3332]{Trevor Bowen}
\affiliation{Space Sciences Laboratory, University of California, Berkeley, CA 94720-7450, USA}

\author[0000-0001-7222-3869]{Nooshin Davis}
\affiliation{Space Science Center and Department of Physics, University of New Hampshire, Durham, NH 03824, USA}

\author[0000-0003-4177-3328]{B. D. G. Chandran}
\affiliation{Space Science Center and Department of Physics, University of New Hampshire, Durham, NH 03824, USA}

\author[0000-0002-6276-7771]{Lorenzo Matteini}
\affiliation{Imperial College London, South Kensington Campus, London SW7 2AZ, UK}

\author[0000-0002-7317-8665]{Ning Kang (康宁)}
\affiliation{Department of Atmospheric and Oceanic Sciences, University of California, Los Angeles, CA, USA}

\author[0000-0003-3367-5074]{Xiaofei Shi (石晓霏)}
\affiliation{Department of Earth, Planetary, and Space Sciences, University of California, Los Angeles, CA, USA}

\author[0000-0002-9954-4707]{Jia Huang (黄佳)}
\affiliation{Space Sciences Laboratory, University of California, Berkeley, CA 94720-7450, USA}

\author[0000-0002-1989-3596]{Stuart D. Bale}
\affiliation{Physics Department, University of California, Berkeley, CA 94720-7300, USA}
\affiliation{Space Sciences Laboratory, University of California, Berkeley, CA 94720-7450, USA}

\author[0000-0002-7077-930X]{J. C. Kasper}
\affiliation{BWX Technologies, Inc., Washington DC 20001, USA.}
\affiliation{Climate and Space Sciences and Engineering, University of Michigan, Ann Arbor, MI 48109, USA.}

\author[0000-0001-5030-6030]{Davin E. Larson}
\affiliation{Space Sciences Laboratory, University of California, Berkeley, CA 94720, USA.}

\author[0000-0002-0396-0547]{Roberto Livi}
\affiliation{Space Sciences Laboratory, University of California, Berkeley, CA 94720, USA.}

\author[0000-0002-7287-5098]{P. L. Whittlesey}
\affiliation{Space Sciences Laboratory, University of California, Berkeley, CA 94720, USA.}

\author[0000-0003-0519-6498]{Ali Rahmati}
\affiliation{Space Sciences Laboratory, University of California, Berkeley, CA 94720, USA.}

\author[0000-0002-5699-090X]{Kristoff Paulson}
\affiliation{Smithsonian Astrophysical Observatory, Cambridge, MA 02138 USA.}

\author[0000-0002-7728-0085]{M. Stevens}
\affiliation{Smithsonian Astrophysical Observatory, Cambridge, MA 02138 USA.}

\author[0000-0002-3520-4041]{A. W. Case}
\affiliation{Smithsonian Astrophysical Observatory, Cambridge, MA 02138 USA.}

\author[0000-0002-4401-0943]{Thierry {Dudok de Wit}}
\affil{LPC2E, CNRS and University of Orl\'eans, Orl\'eans, France}


\author[0000-0003-1191-1558]{David M. Malaspina}
\affil{Astrophysical and Planetary Sciences Department, University of Colorado, \\ Boulder, CO, USA}
\affil{Laboratory for Atmospheric and Space Physics, University of Colorado, \\ Boulder, CO, USA}

\author[0000-0002-0675-7907]{ J.W. Bonnell}
\affil{Space Sciences Laboratory, University of California, Berkeley, CA 94720-7450, USA}

\author[0000-0003-0420-3633]{Keith Goetz}
\affiliation{School of Physics and Astronomy, University of Minnesota, Minneapolis, MN 55455, USA}

\author[0000-0002-6938-0166]{Peter R. Harvey}
\affil{Space Sciences Laboratory, University of California, Berkeley, CA 94720-7450, USA}

\author[0000-0003-3112-4201]{Robert J. MacDowall}
\affil{Solar System Exploration Division, NASA/Goddard Space Flight Center, Greenbelt, MD, 20771}

\begin{abstract}

The trace magnetic power spectrum in the solar wind is known to be characterized by a double power law at scales much larger than the proton gyro-radius, with flatter spectral exponents close to -1 found at the lower 
frequencies below an inertial range with indices closer to $[-1.5,-1.6]$. The origin of the $1/f$ range is still under debate. In this study, we selected 109 magnetically incompressible solar wind intervals ($\delta |\boldsymbol B|/|\boldsymbol B| \ll 1$) from Parker Solar Probe encounters 1 to 13 which display such double power laws, with the aim of understanding the statistics and radial evolution of the low frequency power spectral exponents from Alfv\'en point up to 0.3 AU.
New observations from closer to the sun show that in the low frequency range solar wind turbulence can display spectra much shallower than $1/f$, evolving asymptotically to $1/f$ as advection time increases, indicating a dynamic origin for the $1/f$ range formation. We discuss the implications of this result on the Matteini et al. (2018) conjecture for the $1/f$ origin as well as example spectra displaying a triple power law consistent with the model proposed by Chandran et al. (2018), supporting the dynamic role of parametric decay in the young solar wind. Our results provide new constraints on the origin of the $1/f$ spectrum and further show the possibility of the coexistence of multiple formation mechanisms.

\end{abstract}


\section{Introduction} \label{sec:intro}
The trace magnetic power spectrum (PSD) in the solar wind is often characterized by a double power law from intermediate to large scales, with power spectral exponents in the inertial range around -5/3 (or closer to -3/2 in the inner heliosphere) and -1 at larger scales, respectively. This double power law is usually found in the fast wind \citep{bavassano_statistical_1982,denskat_statistical_1982,burlaga_radial_1984}, and recently in very long intervals of slow wind \citep{bruno_low-frequency_2019}, and also in extremely long interval without regard to wind speed \citep{matthaeus_low-frequency_1986}. The low frequency (or large scale in configuration space, converted with modified Taylor Hypothesis \cite{taylor_spectrum_1938,perez_applicability_2021}) range of the spectrum has been considered as the energy reservoir that facilitates the turbulence cascade in the solar wind. The origin and formation mechanism of the $1/f$ range of the PSD is still not well-understood and under active debate (see e.g. \cite{matthaeus_low-frequency_1986, velli_turbulent_1989,dmitruk_low-frequency_2007, bemporad_low-frequency_2008, verdini_origin_2012, matteini_1_2018,chandran_parametric_2018,magyar_phase_2022}). It is worth noting that the $1/f$ range of the spectrum is indicative of a scale-independent fluctuation energy distribution, i.e. equipartition of energy over all scales \citep{keshner_1f_1982}. Moreover, $1/f$ spectrum is also seen in plasma density fluctuations from {\it Ulysses} at similar frequencies, especially in the high latitudes intervals \citep{matthaeus_density_2007}.

The majority of the models proposed for the double power law were built on an {\it a priori} assumption of the existence of $1/f$ spectrum at low frequency. For example \cite{velli_turbulent_1989} proposed that the secondary incoming waves generated by linear coupling of the dominant outgoing waves to the large scale solar wind inhomogeneity could facilitate a quasistationary self-similar cascade, resulting in a $1/k$ scaling. \citep{montroll_1f_1982, matthaeus_low-frequency_1986} suggest that the $1/f$ spectrum results from the superposition of uncorrelated samples of solar surface turbulence that have log-normal distributions of correlation lengths corresponding to a scale-invariant distribution of correlation times over an appropriate range of parameters. The scale invariance originates in the dynamo and manifests in the photospheric magnetic field.

There are however few notable exceptions. For example it has been pointed out by \cite{chandran_parametric_2018} (henceforth denoted as C18) that in Fig 2-2 of \cite{tu_mhd_1995}, the low frequency $z^+$ (Elsasser variables: $z^\pm=\boldsymbol{V}\mp \boldsymbol{B}/\sqrt{\mu_0 \rho}$) spectrum were as shallow as $f^{-0.5}$ in the low frequency range, which C18 referred as `infrared' range. C18 used this as an input to the model and the parametric decay induced inverse cascade of outward propagating Alfv\'en wave would eventually produce a triple power law $z^+$ spectrum with $1/f$ in the middle frequency range between `infrared' range and the inertial range.

Another example would be the conjecture proposed by \cite{matteini_1_2018} (henceforth denoted as M18) which tried to build a connection between spectral properties and magnetic compressibility. In particular, there's a paradox between an arbitrary power law index and low magnetic compressibility at all scales. Their conjecture shows that, for turbulence with low magnetic compressibility, if the magnetic fluctuations fully saturate over the scales, i.e. $\langle \delta B\rangle = \langle \delta \boldsymbol{B}\rangle\sim |B|=B$, we have $\langle \delta B/B\rangle\sim \langle \delta B\rangle/B \sim 1$. According to the well-known relation which connects the slope in the power spectrum $P(k)\propto k^{\alpha}$, and the exponent in the second order structure function $\delta B^2 \propto l^{-2\beta}$ via: $\alpha = 2\beta - 1$, $-3<\alpha<-1$ (see e.g. \citep{monin_statistical_1987}, and here we translate the frequency $f$ in to wavenumber $k$ with Taylor Hypothesis, and scale $l\sim 1/k$), we see that when $\langle \delta B/B\rangle\sim 1$ for sufficiently large scales (Note that $\langle \delta B/B\rangle$ is a function of temporal increment $\tau$ or scales $l$ in configuration space.), we have $\beta \sim 0$, and hence $\alpha \gtrsim -1$, for the same scale range. Consequently, $1/f$ is the steepest possible realization of power spectrum when magnetic fluctuations fully saturate to the $\langle \delta B/B\rangle \sim 1$ state. In addition, their result leads to a straightforward connection between the scale $l_0$ at which magnetic field fluctuations fully saturate (i.e. $\langle \delta B/B \rangle\sim 1$) and the spectral breakpoint $f_{B}$ of the double power law in PSD. M18 argued that closer to the sun, unless the $1/f$ range is formed in the solar corona and advected outwards preserving its shape, it should gradually disappear moving closer to the sun where $\delta B/B < 1$.

Previous observations with {\it Ulysses}, {\it Helios}, {\it WIND} have only explored the heliosphere beyond 0.3 AU. It is well known that due to solar wind expansion, $B$ decreases like $R^{-2}$ radially per conservation of magnetic flux (or Parker Spiral \citep{parker_dynamics_1958}, because the field is mostly radial close to the sun) and $\delta B$ decreases like $R^{-1.5}$ per WKB theory (see e.g. \cite{whang_alfven_1973,heinemann_non-wkb_1980,velli_waves_1991,tu_mhd_1995,huang_conservation_2022}), where $R$ is heliocentric distance. And hence $\langle \delta B/B\rangle$ is expected to increase radially as $R^{0.5}$ from Alfv\'en critical point (where $\delta B^2, \delta V^2 \propto V_{sw} V_A/(V_A+V_{sw})^2$ reaches its absolute maximum per WKB theory. However, the peak is expected to locate inside the Alfv\'en point due to non-WKB effects, see e.g. \cite{cranmer_generation_2005, verdini_alfven_2007,chen_evolution_2020,bandyopadhyay_sub-alfvenic_2022}) until it reaches its full saturation at $1$ around 0.3 AU. Therefore, closer to the sun, based on the WKB prediction, we expect to see magnetically incompressible intervals with partial saturation ($\langle\delta B/B\rangle \sim const, const < 1$), or without saturation ($\langle \delta B/B\rangle$ keeps increasing over scales but smaller than $1$). It is hence important to validate the connection between $1/f$ spectrum and low magnetic compressibility with solar wind closer to the sun where full magnetic fluctuations saturation is not achieved.


As a consequence, with new observations much closer to the sun from Parker Solar Probe (PSP) \citep{fox_solar_2016}, there are two major questions to be investigated: 1. What are the statistics and radial evolution of low frequency power spectral exponents; 2. With the new evidence from closer to the sun, what new constraints can be applied to the existing models of $1/f$ spectrum. To answer these questions we perform a systematic search for magnetically incompressible ($\delta |\boldsymbol B|/|\boldsymbol B| \ll 1$) solar wind intervals that are characterized by a double power law with PSP data from Encounter 1 to Encounter 13. The rest of the paper is organized as follows: In section 2, we explain the data selection and analysis procedure; In section 3, we show the main results; In section 4, we discuss the implications from new observations on \cite{matteini_1_2018} and \cite{chandran_parametric_2018}; In section 5, we conclude and summarize the main results.


\begin{figure*}
     \centering
     \includegraphics[width=0.98\textwidth]{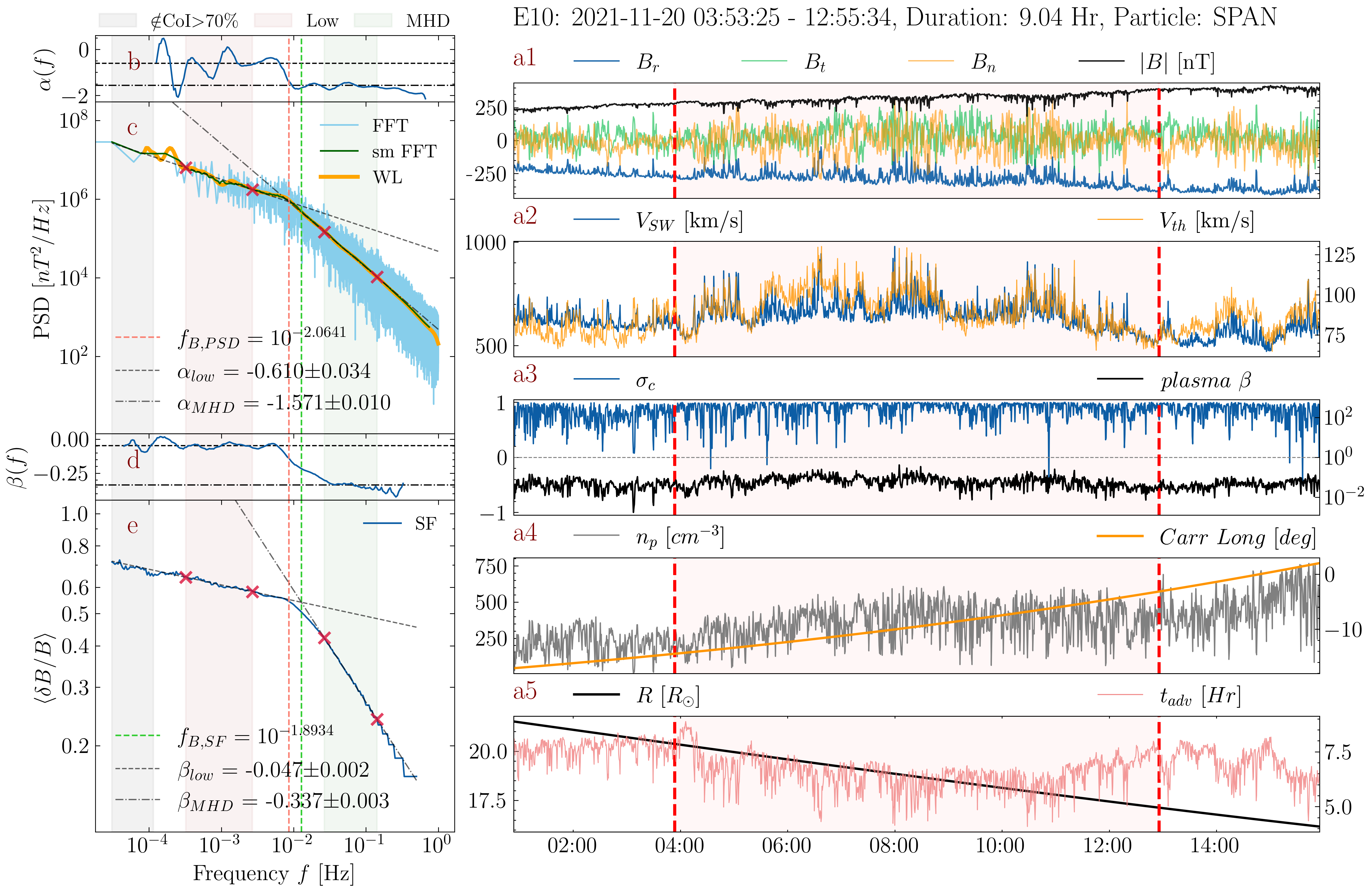}
     \caption{
          Example of a magnetically incompressible solar wind interval from PSP Encounter 10, close to the perihelion. The particle data of this interval are provided by SPAN \citep{livi_solar_2022}. 
          (a1-a5) show time series of magnetic field $B_{rtn}$ in RTN coordinates; solar wind speed $V_{SW}$ and solar wind thermal speed $V_{th}$; cross helicity $\sigma_c$ and plasma beta $\beta$; proton density $n_p$ and Carrington longitude; and heliocentric distance $R$ in solar radii $R_{\odot}$ and advection time $t_{adv} = R/V_{SW}$, respectively. The selected interval is indicated with the red shaded area enclosed by two red dashed lines.
          (b) Power law fitting index $\alpha$ as a function of frequency $f$ of trace PSD.
          (c) Fast Fourier Transformation (FFT) is plotted in blue as background, smoothed FFT PSD is the green line, and Morlet Wavelet Transformation (WL) PSD is the orange line. The two pairs of red crosses accompanied with red and green shaded areas indicate the ranges of frequency over which the power-law fits are applied, and the two fitted lines are shown in dashed and dashed-dot lines. The gray shaded area is the frequency range where more than 70\% of points fall out of the Cone of Influence (CoI) of wavelet transformation. The red dashed line indicates the intersect of two fitted lines, interpreted as the low frequency break point.
          (d) Power law fitting index $\beta$ as a function of frequency $f$ of the normalized structure function SF.
          (e) Normalized first order structure function $\langle \delta B/B\rangle$ as a function of frequency $f=1/(2*\Delta t)$, where $\Delta t$ is the temporal increment. Two fits are applied to the same frequency ranges from (c), which produce a similar low frequency break point, indicated by the green dashed line.
     }
     \label{fig:1}
     \end{figure*}

\section{Data and Analysis Procedures} \label{sec:Data and Analysis Procedures}
In order to study the low frequency spectral properties of magnetically incompressible solar wind observed by PSP, we performed a systematic search for magnetically incompressible turbulence intervals from E1 (Nov 2018) to E12 (Jun 2022). We mainly use high-resolution vector magnetic field measurements from FIELDS instruments \citep{bale_fields_2016} to calculate the magnetic turbulence spectrum, and proton measurements from SWEAP instruments  \citep{kasper_solar_2016} and Quasi Thermal Noise (QTN) electron density measurements \citep{moncuquet_first_2020,pulupa_solar_2017} to support interval selection.

\subsection{Interval Selection}
Our selection of magnetically incompressible solar wind intervals is based on the evaluation of the scale-dependent magnetic compressibility $\eta_B(T)$:

\begin{eqnarray}
     \eta_B(T) = 
     \left\langle\frac{|B| - \langle |B| \rangle _{T}}{\langle |B| \rangle}\right\rangle
\end{eqnarray}
where $\langle \rangle _T$ is the ensemble average at scale T, and $\langle\rangle$ is the ensemble average throughout the interval. For all intervals, we ensure that $\eta_B$ is smaller than 0.1 for any given scale $T$.
Moreover, due to the rapid movement of the spacecraft around perihelia, careful selection is needed to ensure that the spacecraft stays within the same type of solar wind over the selected interval. Therefore secondary parameters, including solar wind speed $V_{sw}=|\boldsymbol{V}|$, thermal speed $V_{th}=\sqrt{2 k_{B} T/m_p}$, normalized cross helicity $\sigma_c=({z^{+}}^2-{z^{-}}^2)/({z^{+}}^2+{z^{-}}^2)$, plasma $\beta=2\mu_0 P/B^2$, proton density $n_p$, Carrington longitude, Heliocentric distance $R$, and advection time $t_{adv}=R/V_{sw}$ are examined to differentiate different solar types (see Figure \ref{fig:1} panels a1-a5). With our selection criteria, we ended up with 109 non-overlapping magnetically incompressible solar wind intervals observed by PSP from E1 to E13, with total 1500 hours worth of data (see Figure \ref{fig:2}c for histogram of interval lengths). Note that the jets in the solar wind speed in panel a2, which are accompanied with partial or total reversal of $B_r$ in panel a1 are known as switchbacks, and have been studied by numerous recent studies (see e.g. \cite{bale_highly_2019,dudok_de_wit_switchbacks_2020,tenerani_magnetic_2020,farrell_magnetic_2020,mozer_time_2020,woolley_proton_2020,bourouaine_turbulence_2020,martinovic_multiscale_2021,larosa_switchbacks_2021,tenerani_evolution_2021,hernandez_impact_2021,laker_statistical_2021,meng_analysis_2022,shi_patches_2022,huang_structure_2023} and references therein). They are known as large amplitude or spherically polarized Alfv\'en wave \citep{mallet_evolution_2021,mallet_exact_2021} with almost constant magnetic modulus. 

\subsection{Diagnostics: Magnetic Trace Power Spectrum and Structure Function}\label{sec:Diagnostics: Magnetic Spectrum and Structure Function}
For each interval, the following two diagnostics are calculated: trace power spectrum density (PSD) and normalized first-order structure function (SF) for vector magnetic field. To produce a reliable low frequency turbulence spectrum, for each interval, we calculate the trace power spectrum density (PSD) with both Morlet Wavelet Transformation (WL) and Fast Fourier Transformation (FFT). For FFT, we smooth the spectrum by averaging over a sliding window of a factor of 2 in the frequency domain (sm-FFT). The smooth spectra calculated with the two methods (WL and sm-FFT) generally overlap with each other perfectly in the high frequency range but gradually deviate from each other at the low frequency end. Therefore, we keep the spectrum up to the frequency where more than 70\% of points fall out of the Cone of Influence (CoI) of the WL spectrum, and we trust only the frequency range where WL overlap with sm-FFT. We calculate the power law fit index $\alpha$ by fitting on the WL PSD unless specified otherwise, and henceforth referred simply as PSD.

We follow the steps described in M18 to calculate the normalized first-order structure function for vector magnetic field. We apply 200 logarithmically spaced lags $\Delta t$ as temporal increment of vector magnetic field $\delta \boldsymbol{B}(t, \Delta t) = \boldsymbol B(t)-\boldsymbol B(t+\Delta t)$ within the range ($1s$, $length\ of \ interval /2$). The normalized first-order structure function is then calculated by:
\begin{eqnarray}
     SF(\Delta t) = SF_{norm}(\Delta t) = \left\langle \frac{|\delta \boldsymbol B(t, \Delta t)|}{B(t, \Delta t) }\right\rangle _{t}
\end{eqnarray}
where we denote $|\delta \boldsymbol B|$ as $\delta B$ and $|\boldsymbol{B}|$ as $B$, and $\langle \rangle _{t}$ is average with regard to $t$ for scale $\Delta t$. Note that the intervals we select are magnetically incompressible, and hence $B$ can be considered as scale-independent and can be taken out of the averaging window $\langle \rangle _t$. The resultant $SF(\Delta t)$ in the temporal domain is finally converted to $SF(f)$ in the frequency domain via $f = 1/(2\cdot\Delta t)$ because a half period in the absolute fluctuation corresponds to the full period in the non-absolute one.

To extract spectral information from PSD, we apply power law fit on PSD in both low and MHD inertial frequency ranges where the spectrum stabilize visually (red and green shaded area in Figure \ref{fig:1}c), and we obtain two power law fits with indices $\alpha_{Low}$ and $\alpha_{MHD}$ (black dashed line and dashed dotted line in Figure \ref{fig:1}c). The intersect of the two fit lines is interpreted as the low frequency break point ($f_{B,PSD}$, red dashed line). To compare the spectral properties of PSD and SF, we apply power law fit in the same frequency ranges to SF (extended shaded area in Figure \ref{fig:1}e), and the intersect of two fit lines is also interpreted as the low frequency break point ($f_{B,SF}$, green dashed line). Clearly the two break frequencies $f_{B,PSD}$ and $f_{B,SF}$ are very close to each other, consistent with M18. The small deviation may be attributed to spectral leakage of the structure function, which leads to a spectrum smoother than the wavelet or PSD result. For completeness, two moving fits $\alpha(f)$ and $\beta(f)$ with window size of 1/3 decade on PSD and SF are shown in Figure \ref{fig:1}b and \ref{fig:1}d.

\section{Results} \label{sec:Statistical Results}

\begin{figure}
     \centering
     \includegraphics[width=0.47\textwidth]{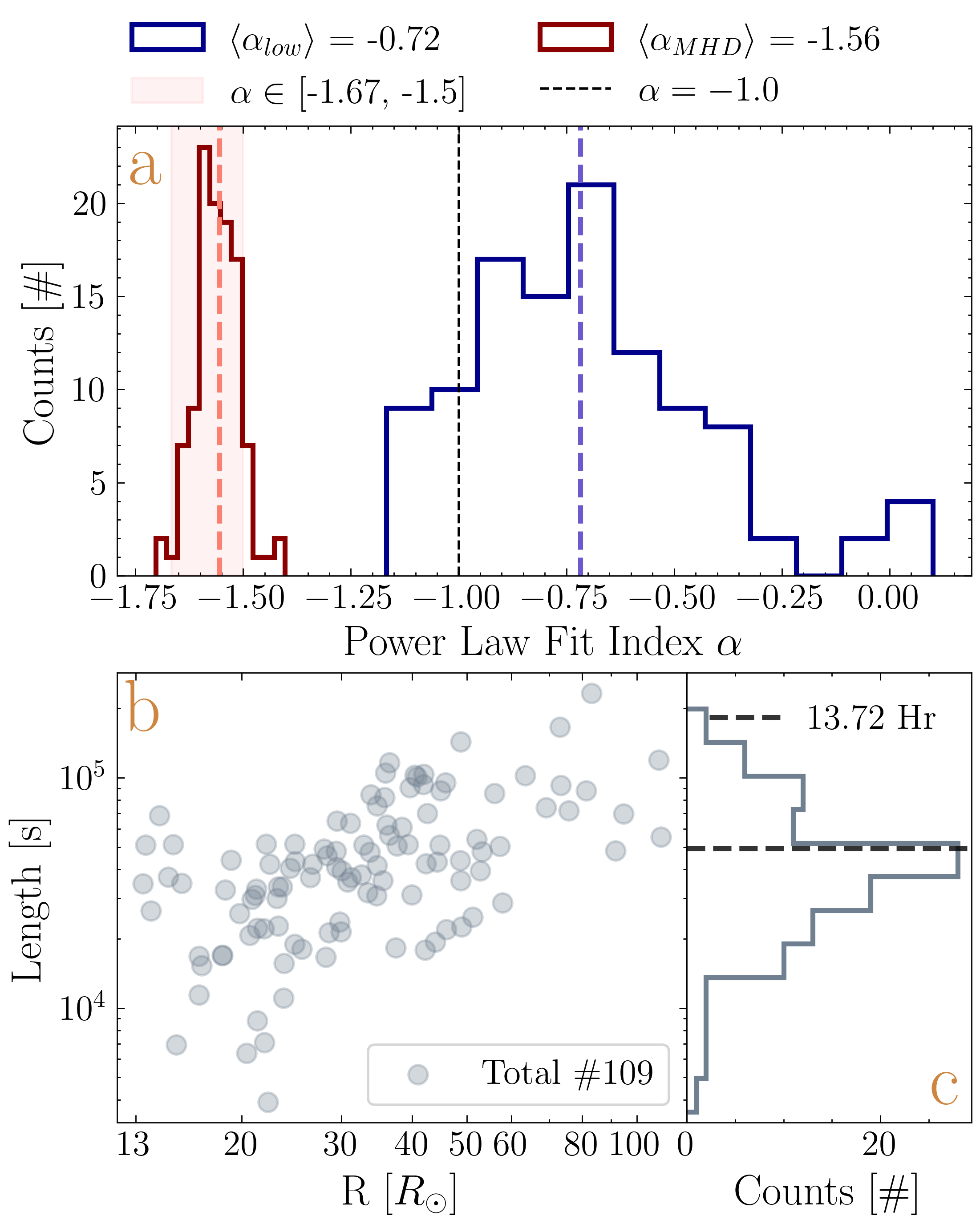}
     \caption{
          (a) Histograms of power law fit index $\alpha$ of low frequency (blue) and MHD frequency (red) ranges. The averaged index values are indicated with corresponding dashed lines and are shown in the legends. The expected power spectral exponent value range from various phenomenologies is indicated with red shaded area.
          (b) Statistics of length and mean radial distance of the selected 109 intervals.
          (c) Histogram of interval lengths. The average length is 13.72 hours (shown as legend with accompanied with dashed line), with minimum of 1.08 hours and maximum of 64.70 hours.
     }
     \label{fig:2}
\end{figure}

\subsection{Statistics of Power Spectral Exponents}\label{subsec:Statistics of Power Spectral Exponents}
For the 109 intervals, power law fit is applied to both high frequency (MHD/inertial) and low frequency ranges with the method described in section \ref{sec:Diagnostics: Magnetic Spectrum and Structure Function}. The primary statistical results are presented in Figure \ref{fig:2}a, in a style similar to Figure 6 in \cite{bruno_low-frequency_2019}. The histogram of power law fit index in the MHD inertial range $\alpha_{MHD}$ is shown in dark red. Most of the intervals have an inertial range spectral exponent that falls within the expected value range [-1.67, -1.5],  predicted by many existing phenomenologies \citep{kolmogorov_local_1941,iroshnikov_turbulence_1964,kraichnan_inertialrange_1965,sridhar_toward_1994, goldreich_magnetohydrodynamic_1997,boldyrev_spectrum_2005, boldyrev_spectrum_2006}, consistent with recent observations (see e.g. \cite{chen_evolution_2020,telloni_evolution_2021,shi_alfvenic_2021,kasper_parker_2021,zank_turbulence_2022,sioulas_magnetic_2023,sioulas_evolution_2023,raouafi_parker_2023}). However, for the low frequency range fit index ($\alpha_{Low}$, dark blue), the spectral exponents have an unexpectedly wide distribution, and the majority of them are larger than $-1$, i.e. the corresponding low frequency spectra are shallower than $1/f$. 


\begin{figure*}
     \centering
     \includegraphics[width=0.95\textwidth]{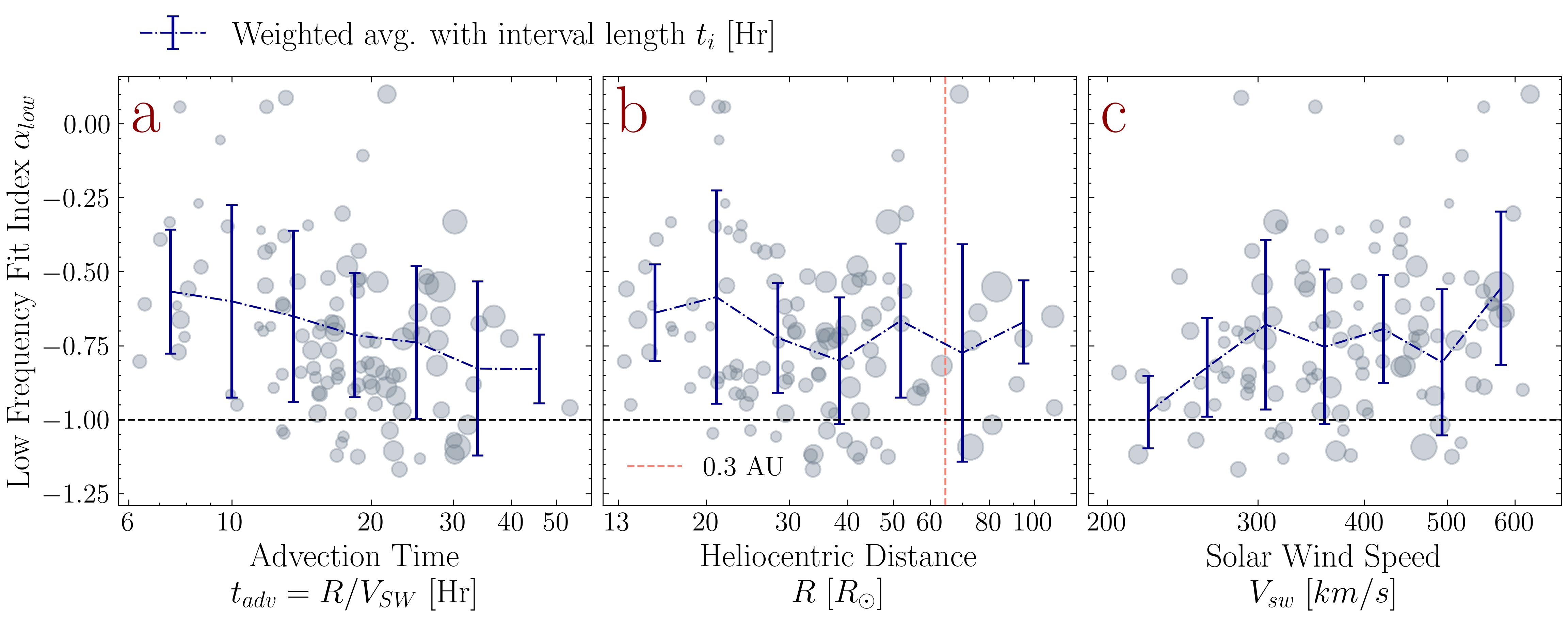}
     \caption{
          (a) Radial evolution of low frequency power law fit index $\alpha_{Low}$, sorted with advection time ($t_{adv} = R/V_{SW}$). The sizes of the circles indicate the interval length (weight), and the line plot is the binned weighted average of $\alpha_{Low}$. The errorbars are one standard error of $\alpha_{Low}$ from each bin.
          (b) Radial evolution of $\alpha_{Low}$ sorted with Heliocentric distance $R$.
          (c) Dependence of $\alpha_{Low}$ on solar wind speed $V_{SW}$.
     }
     \label{fig:3}
\end{figure*}

\subsection{Radial Evolution and Dependence on Solar Wind Speed}\label{subsec:Radial Evolution and Dependence on Solar Wind Speed}
The radial evolution of power law fit index of PSD in the low frequency range $\alpha_{Low}$ is shown in Figure \ref{fig:3}, sorted with advection time $t_{adv}=R/V_{SW}$ in \ref{fig:3}a and heliocentric distance $R$ in \ref{fig:3}b. To display the trend of evolution, a binned average weighted with interval length is plotted on top of the scatters. When we sort the intervals with advection time $t_{adv}$, a clear asymptotic evolution from shallower spectrum towards $1/f$ spectrum is seen. However, when sorted with heliocentric distance $R$, no clear trend is found, and instead $\alpha_{Low}$ is scattered in a wide range of values below 0.3 AU. For completeness, the dependence of $\alpha_{Low}$ on solar wind speed is shown in panel c. For intervals with very low solar wind speed ($\sim 200$ km/s), which are typically observed very close to the sun, $\alpha_{Low}$ are mostly close to -1, i.e. very close to $1/f$ spectrum. For intervals with higher solar wind speed, no obvious trend is observed. Therefore, for the very low speed streams, it is possible that the $1/f$ spectrum originated from the solar corona (see e.g. \cite{matthaeus_low-frequency_1986}). 

However, substantial solar wind acceleration has been observed below 0.3 AU (see e.g. \cite{shi_alfvenic_2021,sioulas_magnetic_2022,shi_acceleration_2022}). In other words, the solar wind speed is a radially varying parameter even for the same stream, and hence the dependence of $\alpha_{Low}$ on solar wind speed should be taken with caution. Moreover, it should be also noted that the actual advection time is an integrated quantity, and thus the $t_{adv}$ used here can be merely considered as a proxy to the real advection time. And the interval length systematically grows with advection time and heliocentric distance, and hence $\alpha_{low}$ is acquired from lowering frequency range with increasing $t_{adv}$ and $R$. The radial evolution of $\alpha_{low}$ could be associated with the change in fitting frequency range.

\begin{figure}
     \centering
     \includegraphics[width=0.47\textwidth]{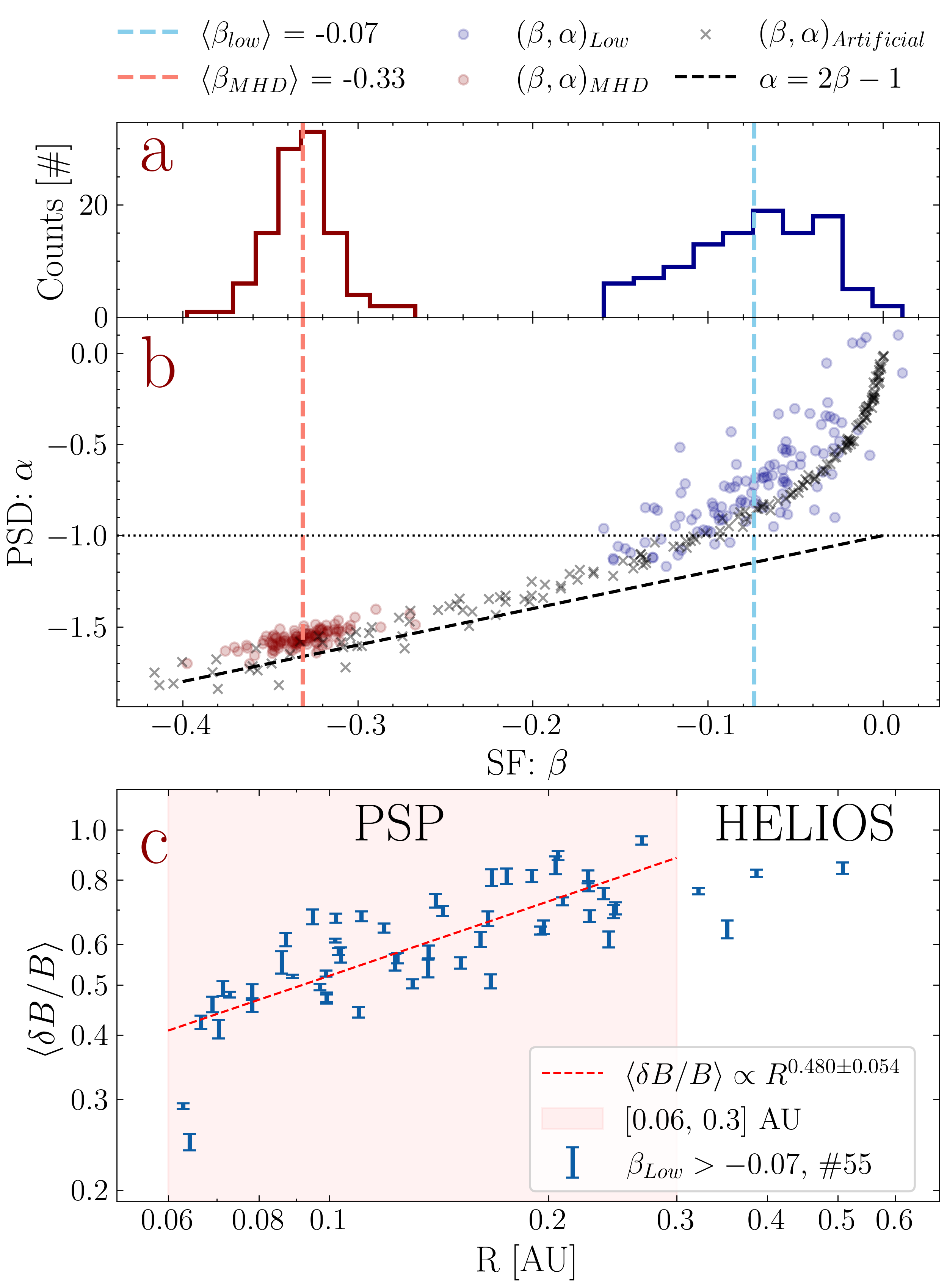}
     \caption{
          (a) Histogram of power law fit index $\beta$ of the normalized first order structure function SF from both low (blue) and MHD (red) frequency ranges. The averages are indicated with corresponding dashed lines.
          (b) Scatters of power law fit index pairs ($\alpha$,$\beta$) of trace PSD and SF from both low (blue) and MHD (red) frequency ranges. The ($\alpha$,$\beta$) pairs from artificial data are also shown with black crosses, and the theoretical relation $\alpha=2\beta-1$ is indicated with black dashed line.
          (c) Dependence of $\langle \delta B/B\rangle$ from the saturated intervals ($\beta_{Low} > -0.07$) on Heliocentric distance $R$. The value ranges of $\langle \delta B/B\rangle$ in the low frequency range are indicated with errorbars. The newly explored Heliosphere radial range ($R<0.3\ AU$) is indicated with red shaded area.
     }
     \label{fig:4}
\end{figure}

\section{Discussions}
\subsection{Implications on Matteini2018}
M18 tried to build a connection between low magnetic compressibility and $1/f$ spectrum via the well-known relation: $\alpha=2\beta-1,\ -1>\alpha>-3$, where $\alpha$ is the spectral exponent of the PSD, and $\beta$ is the power law index of the first order structure function (here expressed in terms of frequency). Due to low magnetic compressibility, $\langle \delta B/B\rangle$ can be considered as first order structure function $\delta B$ normalized by its own constant modulus. Therefore the fully saturated magnetic fluctuations, i.e. $\delta B/B \sim 1$, can be translated to $\beta \sim 0$, and therefore $\alpha \gtrsim -1$. In other words, $1/f$ is the steepest possible realization of the low frequency spectrum when magnetic field fluctuations fully saturates. It has also been proposed by M18 that ``unless the 1/f range is formed in the corona and just advected in interplanetary space preserving its shape, it should gradually disappear, moving closer to the Sun where $\delta B/B < 1$." 

It is therefore interesting to examine the behaviors of $\alpha_{Low}$ and $\beta_{Low}$ closer to the sun where $\delta B/B <1$. There are roughly two possible behaviors of $\langle \delta B/B \rangle$ in the low frequency range: 1. Partial saturation, i.e. $\delta B/B \sim const$, $\beta \sim 0$; 2. No saturation, $\beta < 0$. Figure \ref{fig:4}a shows the histogram of the $\beta_{Low}$ and $\beta_{MHD}$ in both low and MHD frequency ranges. The histogram of $\beta_{Low}$ clearly shows that substantial portion of intervals have no saturation ($\beta \lesssim -0.1$). However, the scatter plot of ($\beta_{Low}$, $\alpha_{Low}$) in Figure \ref{fig:4}b shows that the low frequency PSD of these intervals are close to $1/f$. In fact for those intervals with partial saturation ($\beta_{Low} \gtrsim -0.1$), the scatter plot shows that the corresponding $\alpha_{Low}$ are much greater than $-1$, i.e. the low frequency PSD are much shallower than $1/f$. 

Figure \ref{fig:4}b further reveals that in the inertial range, the observed ($\beta$, $\alpha$) pairs follow the relation $\alpha = 2\beta-1$ pretty well (black dashed line), with a slight systematic shift towards shallower $\alpha$, but in the low frequency range deviate significantly. To understand such deviation, we generate 1-D artificial time series with different power spectral exponent $\alpha_0$. For each $\alpha_0$, we generate an artificial time series $A(t;\alpha_0)$ formed with sinusoidal fluctuations with random phases and amplitudes following power law:
\begin{eqnarray}
     A(t;\alpha_0) = \sum_{i}^N \omega_i^{(\alpha_0+1)/2} \sin[2\pi(\omega_i t + \phi_i)]
\end{eqnarray}
where $\omega_i$ is $i^{th}$ frequency log-linearly spaced from $10^{-5}$ Hz to $1$ Hz, $\phi_i$ is random phase in range of $(0, 2\pi)$, and $N$ is set to 10000. For each $A(t;\alpha_0)$, we calculate the power law fitting index for the PSD ($\alpha$) and SF ($\beta$) from the same frequency range $(10^{-2.5}, 10^{-1})$ Hz. 

The scatter plot of $(\alpha, \beta)_{Artificial}$ pairs are shown in Figure \ref{fig:4}b as gray crosses. $(\alpha, \beta)_{Artificial}$ catch well the trend of the observed $(\alpha, \beta)$ pairs, especially in the MHD frequency range, and also deviate significantly from the relation $\alpha=2\beta-1$. Such systematic deviation is possibly due to the spectral leakage effect, because contrary to FFT and wavelet transformation, the structure function has a broad-band response. Therefore, cautions should be taken when one is trying to relate $1/f$ spectrum with magnetic fluctuations saturation (regardless of full or partial). For completeness, shown in Figure \ref{fig:4}c, for those 55 intervals with partially saturated magnetic fluctuations (here defined as $\beta > -0.07$), we see the radial evolution of saturation values $\langle \delta B/B \rangle$ follows well with the WKB prediction of $R^{0.5}$. 

In summary, one should be careful when relating magnetic fluctuations saturation ($\langle \delta B/B\rangle\sim const$) to $1/f$ spectrum, because of the spectral leakage effect. And closer to the sun where $\delta B/B < 1$, we see low frequency spectra much shallower than $1/f$ with partially saturated magnetic fluctuations. One good example has  already been shown in Figure \ref{fig:1}. Therefore, according to M18, it is possible that some of these low frequency range spectra originated from the solar corona (see also \cite{matthaeus_evaluation_1982,matthaeus_density_2007,bemporad_low-frequency_2008} for the coronal origin of $1/f$ spectrum.)

\begin{figure}
     \centering
     \includegraphics[width=0.47\textwidth]{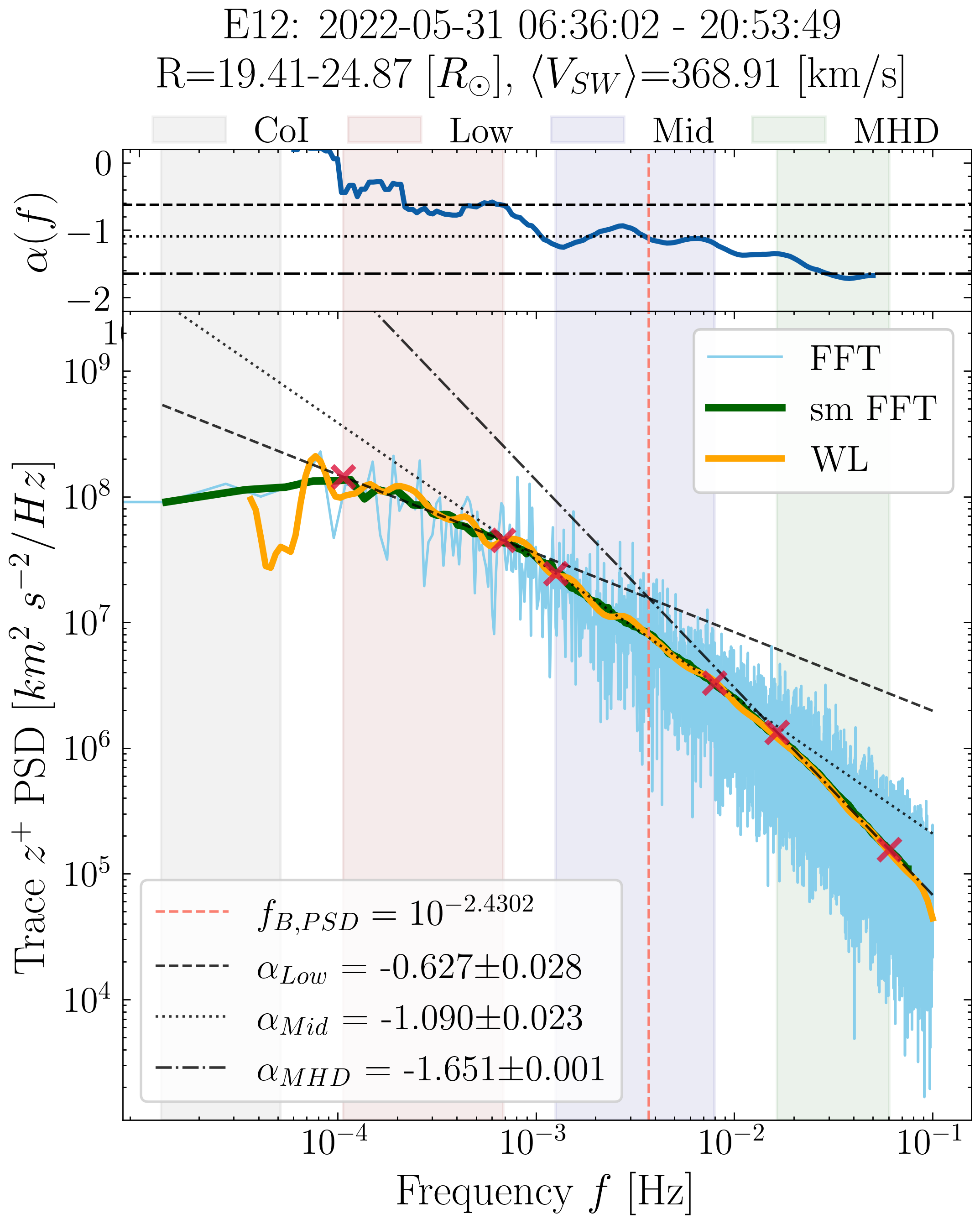}
     \caption{
          Example of C18 type of trace $z^+$ spectrum. The FFT spectrum is shown with blue line as background, the smoothed FFT is the green line, and the trace Morlet Wavelet (WL) PSD is the orange line. The gray shaded area indicates that more than 70\% of points fall out of the Cone of Influence (CoI) of wavelet transformation. The power law fit is conducted on the smoothed FFT, and the moving window fit $\alpha(f)$ (window sized 1/3 decade) is shown in the overhead panel. The trace PSD is separated into three distinct frequency ranges (Low, Mid, MHD), each possesses a quasi-steady power law range, indicated with red (Low), blue (Mid), and green (MHD) shaded areas. Power law fits are applied to each of the ranges and the fit indices $\alpha_{Low/Mid/MHD}$ are shown in the legends. A "low frequency break" is found as the intersect between the power law fits from low frequency and MHD inertial range, indicated with orange dashed line.
     }
     \label{fig:5}
\end{figure}

\subsection{Implications on Chandran2018}
C18 proposed that the $1/f$ spectrum is formed by inverse cascade of Alfv\'en waves facilitated by parametric decay within the context of weak-turbulence theory. The author initialize the simulation with primarily outward propagating Alfv\'en waves, i.e. $e^{+} \gg e^{-}$, where $e^{\pm}$ are the frequency ($f$) spectra of Alfv\'en waves propagating in opposite directions along the magnetic field lines. If the initial $e^{+}$ has a peak frequency $f_0$ (where $f e^+$ is maximized), and an `infrared' scaling $f_p$ at smaller $f$ with $-1<p<1$, then $e^+$ acquires an $f^{-1}$ scaling throughout a range of frequencies that spreads out in both directions from $f_0$. The final state of their model evolution is a triple power law with `infrared' scaling in the low frequency, $1/f$ in the intermediate frequency range, and $f^{-2}$ scaling in the inertial range (the inertial range is initialized with the critical-balanced parallel spectrum of $k_{\parallel}^{-2}$, see e.g. \cite{goldreich_toward_1995,podesta_dependence_2009,forman_detailed_2011}).

Therefore it is interesting to explore the magnetically incompressible (yet weakly compressible, plasma $\beta \ll 1$, and $\delta \vec B \sim |B|$ to allow parametric decay \cite{fu_parametric_2018}), Alfv\'enic turbulence (primarily outward propagating Alfv\'en waves, i.e. $|\sigma_c| \sim 1$) intervals close to the sun, to see if there exists the triple power law $z^{+}$ spectrum. Among the 109 intervals, there are 4 intervals displaying clear triple power law in trace $z^{+}$ PSD with stable intermediate frequency range. One of the best examples is shown in Figure \ref{fig:5}, which is an interval from E12, with $\langle\sigma_c\rangle=0.82$. Note that due to the low resolution particle data, the time series is resampled to 5s, and hence the WL spectrum is very wavy. To acquire a better moving fit $\alpha(f)$ profile, the moving power law fit with window size of 1/3 decade is applied to the smoothed FFT PSD (green line).

In the low frequency range, we see a stable `infrared' spectrum with $f^{-0.627}$. In the intermediate range, the spectrum is very close to $1/f$. And in the inertial range, the scaling (can be converted to configuration space with Taylor Hypothesis, \cite{taylor_relaxation_1986,perez_how_2021}) is agreeable with the perpendicular spectrum of $k_{\perp}^{-5/3}$ from the critical balance theory, which would mask the steeper parallel spectrum because we are observing the trace $z^+$ PSD. This is also supported by \cite{cuesta_isotropization_2022,sioulas_evolution_2023}, which has shown that when considering turbulence anisotropy, for fast streams ($V_{sw} \gtrsim 400 km/s$) close to the sun, the low frequency spectrum is dominated by parallel fluctuations, and in the inertial range, the perpendicular spectrum has a $f^{-5/3}$ scaling. This is also consistent with the simulation by \cite{verdini_origin_2012}, where at higher frequencies the steep parallel spectral slope (-2) is masked by the more energetic perpendicular spectrum (slope -5/3). In summary, this triple power law is consistent with the simulation result from C18, and further provide evidence for the presence of parametric decay instability in the solar wind (see e.g. \cite{matteini_parametric_2010,tenerani_parametric_2013,zanna_parametric_2015,tenerani_parametric_2017,shi_parametric_2017,bowen_density_2018,reville_parametric_2018}).

It should also also noted that due to the existence of the triple power law, the concept of `low frequency break point' is now questionable. For example the orange dashed line in Figure \ref{fig:5} is obtained as the intersect between the low frequency `infrared' spectrum and the MHD spectrum, and interpreted as the low frequency break point $f_{B,PSD}$. However, any frequencies in between these two ranges should be considered as the `break frequency'. Therefore, the low frequency break point obtained with this method would have a unacceptably large uncertainty. For this reason, we have not shown any statistics of $f_{B, PSD}$ in this study.

\section{Conclusions and Summary}
In this letter we have selected 109 magnetically incompressible intervals with total 1500 hours worth of data from Parker Solar Probe encounter 1 to 13. All of the intervals display double power law in the intermediate to large scales in their trace magnetic power spectrum density (PSD). Traditionally, the double power law in PSD is characterized by power indices $-5/3$ and $-1$ at the two scale ranges respectively, and there are many models in the literatures to explain the origin of the $1/f$ range (e.g. \cite{matthaeus_low-frequency_1986, velli_turbulent_1989,dmitruk_low-frequency_2007, verdini_origin_2012, matteini_1_2018,chandran_parametric_2018,magyar_phase_2022}). Previous observations from {\it Ulysses, Helios, WIND} have only explored the heliosphere beyond 0.3 AU. In this study, we aim to use the latest observations from PSP to provide constraints on the origin of the $1/f$ range.

From the statistics of the 109 intervals, we found that within 0.3 AU, the majority the intervals display spectra that are much shallower than $1/f$ in the low frequency range. And as advection time $t_{adv} = R/V_{SW}$ increases, the low frequency power law fit index $\alpha_{Low}$ asymptotically approach $-1$. This suggests a dynamical formation of such range from the Alfv\'en point up to 0.3 AU. However when sorted with heliocentric distance, no obvious evolution of $\alpha_{Low}$ is observed. Moreover, for those extremely slow solar wind streams which were observed very close to sun, the low frequency spectra scaling are very close to $1/f$. Therefore we can not rule out the possibility that the $1/f$ originated from the corona and are advected out \citep{matthaeus_low-frequency_1986}. Moreover, it should be noted that such $1/f$ scaling is also present at the density spectrum from {\it Ulysses} observations \citep{matthaeus_density_2007}, which appeared to have latitudinal dependence at 1AU. At solar mid-latitudes, such scaling matches reasonably well with the longitudinal spectra of photospheric magnetic field \citep{nakagawa_dynamics_1974}.

Unlike other models, \cite{matteini_1_2018} and \cite{chandran_parametric_2018} allow the low frequency spectrum to be shallower than $1/f$. In fact, as has been pointed out by \cite{chandran_parametric_2018}, it has been reported by \cite{tu_mhd_1995} that in the low frequency range, $z^+$ have spectrum as shallow as $f^{-0.5}$. The new observations provide evidence inconsistent with the conjecture by \cite{matteini_1_2018} because closer to the sun where $\delta B/B<1$ the low frequency spectrum does not gradually disappear, but instead they are omnipresent. On the other hand, some Alfv\'enic intervals display clear triple power law in the $z^+$ spectra with $1/f$ scaling in the intermediate frequency range, and therefore provide some evidence for the model from \cite{chandran_parametric_2018}.

The new observations from Parker Solar Probe encounter 1 to 13 provide abundant new evidence for the origin of the low frequency trace magnetic $1/f$ spectrum. Unfortunately, up to this point, the existing models have had difficulty covering all of the observed behaviors of the $1/f$ scaling in trace magnetic PSD. In fact, it is not even appropriate to call this part of the spectrum `$1/f$' spectrum because of the omnipresent shallower spectra observed closer to the sun. However, based on this study, the new evidence indicates that for different intervals from different solar wind conditions, the low frequency spectrum might have different formation mechanisms. Therefore we need to accumulate more observations from future PSP orbits to obtain a clearer picture of the low frequency turbulence spectrum.



\vspace{12pt}
The authors acknowledge the following open source packages: \cite{harris_array_2020,virtanen_scipy_2020,hunter_matplotlib_2007,lam_numba_2015,team_pandas-devpandas_2023,angelopoulos_space_2019}. This research was funded in part by the FIELDS experiment on the Parker Solar Probe spacecraft, designed and developed under NASA contract UCB \#00010350/NASA NNN06AA01C, and the NASA Parker Solar Probe Observatory Scientist grant NASA NNX15AF34G. M. Velli acknowledges support from ISSI via the J. Geiss fellowship.

\bibliography{sample631}

\begin{thebibliography}{}
\expandafter\ifx\csname natexlab\endcsname\relax\def\natexlab#1{#1}\fi
\providecommand{\url}[1]{\href{#1}{#1}}
\providecommand{\dodoi}[1]{doi:~\href{http://doi.org/#1}{\nolinkurl{#1}}}
\providecommand{\doeprint}[1]{\href{http://ascl.net/#1}{\nolinkurl{http://ascl.net/#1}}}
\providecommand{\doarXiv}[1]{\href{https://arxiv.org/abs/#1}{\nolinkurl{https://arxiv.org/abs/#1}}}

\bibitem[{Angelopoulos {et~al.}(2019)Angelopoulos, Cruce, Drozdov, Grimes,
  Hatzigeorgiu, King, Larson, Lewis, McTiernan, Roberts, Russell, Hori,
  Kasahara, Kumamoto, Matsuoka, Miyashita, Miyoshi, Shinohara, Teramoto, Faden,
  Halford, McCarthy, Millan, Sample, Smith, Woodger, Masson, Narock, Asamura,
  Chang, Chiang, Kazama, Keika, Matsuda, Segawa, Seki, Shoji, Tam, Umemura,
  Wang, Wang, Redmon, Rodriguez, Singer, Vandegriff, Abe, Nose, Shinbori,
  Tanaka, UeNo, Andersson, Dunn, Fowler, Halekas, Hara, Harada, Lee, Lillis,
  Mitchell, Argall, Bromund, Burch, Cohen, Galloy, Giles, Jaynes, Le~Contel,
  Oka, Phan, Walsh, Westlake, Wilder, Bale, Livi, Pulupa, Whittlesey, DeWolfe,
  Harter, Lucas, Auster, Bonnell, Cully, Donovan, Ergun, Frey, Jackel, Keiling,
  Korth, McFadden, Nishimura, Plaschke, Robert, Turner, Weygand, Candey,
  Johnson, Kovalick, Liu, McGuire, Breneman, Kersten, \&
  Schroeder}]{angelopoulos_space_2019}
Angelopoulos, V., Cruce, P., Drozdov, A., {et~al.} 2019, Space Science Reviews,
  215, 9, \dodoi{10.1007/s11214-018-0576-4}

\bibitem[{Bale {et~al.}(2016)Bale, Goetz, Harvey, Turin, Bonnell, Dudok~de Wit,
  Ergun, MacDowall, Pulupa, Andre, Bolton, Bougeret, Bowen, Burgess, Cattell,
  Chandran, Chaston, Chen, Choi, Connerney, Cranmer, Diaz-Aguado, Donakowski,
  Drake, Farrell, Fergeau, Fermin, Fischer, Fox, Glaser, Goldstein, Gordon,
  Hanson, Harris, Hayes, Hinze, Hollweg, Horbury, Howard, Hoxie, Jannet,
  Karlsson, Kasper, Kellogg, Kien, Klimchuk, Krasnoselskikh, Krucker, Lynch,
  Maksimovic, Malaspina, Marker, Martin, Martinez-Oliveros, McCauley, McComas,
  McDonald, Meyer-Vernet, Moncuquet, Monson, Mozer, Murphy, Odom, Oliverson,
  Olson, Parker, Pankow, Phan, Quataert, Quinn, Ruplin, Salem, Seitz, Sheppard,
  Siy, Stevens, Summers, Szabo, Timofeeva, Vaivads, Velli, Yehle, Werthimer, \&
  Wygant}]{bale_fields_2016}
Bale, S.~D., Goetz, K., Harvey, P.~R., {et~al.} 2016, {\ss}r, 204, 49,
  \dodoi{10.1007/s11214-016-0244-5}

\bibitem[{Bale {et~al.}(2019)Bale, Badman, Bonnell, Bowen, Burgess, Case,
  Cattell, Chandran, Chaston, Chen, Drake, de~Wit, Eastwood, Ergun, Farrell,
  Fong, Goetz, Goldstein, Goodrich, Harvey, Horbury, Howes, Kasper, Kellogg,
  Klimchuk, Korreck, Krasnoselskikh, Krucker, Laker, Larson, MacDowall,
  Maksimovic, Malaspina, Martinez-Oliveros, McComas, Meyer-Vernet, Moncuquet,
  Mozer, Phan, Pulupa, Raouafi, Salem, Stansby, Stevens, Szabo, Velli, Woolley,
  \& Wygant}]{bale_highly_2019}
Bale, S.~D., Badman, S.~T., Bonnell, J.~W., {et~al.} 2019, Nature, 1,
  \dodoi{10.1038/s41586-019-1818-7}

\bibitem[{Bandyopadhyay {et~al.}(2022)Bandyopadhyay, Matthaeus, McComas,
  Chhiber, Usmanov, Huang, Livi, Larson, Kasper, Case, Stevens, Whittlesey,
  Romeo, Bale, Bonnell, Wit, Goetz, Harvey, MacDowall, Malaspina, \&
  Pulupa}]{bandyopadhyay_sub-alfvenic_2022}
Bandyopadhyay, R., Matthaeus, W.~H., McComas, D.~J., {et~al.} 2022, The
  Astrophysical Journal Letters, 926, L1, \dodoi{10.3847/2041-8213/ac4a5c}

\bibitem[{Bavassano {et~al.}(1982)Bavassano, Dobrowolny, Fanfoni, Mariani, \&
  Ness}]{bavassano_statistical_1982}
Bavassano, B., Dobrowolny, M., Fanfoni, G., Mariani, F., \& Ness, N.~F. 1982,
  Solar Physics, 78, 373, \dodoi{10.1007/BF00151617}

\bibitem[{Bemporad {et~al.}(2008)Bemporad, Matthaeus, \&
  Poletto}]{bemporad_low-frequency_2008}
Bemporad, A., Matthaeus, W.~H., \& Poletto, G. 2008, The Astrophysical Journal,
  677, L137, \dodoi{10.1086/588093}

\bibitem[{Boldyrev(2005)}]{boldyrev_spectrum_2005}
Boldyrev, S. 2005, The Astrophysical Journal, 626, L37, \dodoi{10.1086/431649}

\bibitem[{Boldyrev(2006)}]{boldyrev_spectrum_2006}
---. 2006, Physical Review Letters, 96, 115002,
  \dodoi{10.1103/PhysRevLett.96.115002}

\bibitem[{Bourouaine {et~al.}(2020)Bourouaine, Perez, Klein, Chen,
  Martinovi{\'c}, Bale, Kasper, \& Raouafi}]{bourouaine_turbulence_2020}
Bourouaine, S., Perez, J.~C., Klein, K.~G., {et~al.} 2020,
  {\textbackslash}apjl, 904, L30, \dodoi{10.3847/2041-8213/abbd4a}

\bibitem[{Bowen {et~al.}(2018)Bowen, Badman, Hellinger, \&
  Bale}]{bowen_density_2018}
Bowen, T.~A., Badman, S., Hellinger, P., \& Bale, S.~D. 2018, The Astrophysical
  Journal, 854, L33, \dodoi{10.3847/2041-8213/aaabbe}

\bibitem[{Bruno {et~al.}(2019)Bruno, Telloni, Sorriso-Valvo, Marino, Marco, \&
  D{\textquoteright}Amicis}]{bruno_low-frequency_2019}
Bruno, R., Telloni, D., Sorriso-Valvo, L., {et~al.} 2019, Astronomy \&
  Astrophysics, 627, A96, \dodoi{10.1051/0004-6361/201935841}

\bibitem[{Burlaga \& Goldstein(1984)}]{burlaga_radial_1984}
Burlaga, L.~F., \& Goldstein, M.~L. 1984, Journal of Geophysical Research:
  Space Physics, 89, 6813, \dodoi{10.1029/JA089iA08p06813}

\bibitem[{Chandran(2018)}]{chandran_parametric_2018}
Chandran, B. D.~G. 2018, Journal of Plasma Physics, 84,
  \dodoi{10.1017/S0022377818000016}

\bibitem[{Chen {et~al.}(2020)Chen, Bale, Bonnell, Borovikov, Bowen, Burgess,
  Case, Chandran, Wit, Goetz, Harvey, Kasper, Klein, Korreck, Larson, Livi,
  MacDowall, Malaspina, Mallet, McManus, Moncuquet, Pulupa, Stevens, \&
  Whittlesey}]{chen_evolution_2020}
Chen, C. H.~K., Bale, S.~D., Bonnell, J.~W., {et~al.} 2020, The Astrophysical
  Journal Supplement Series, 246, 53, \dodoi{10.3847/1538-4365/ab60a3}

\bibitem[{Cranmer \& Ballegooijen(2005)}]{cranmer_generation_2005}
Cranmer, S.~R., \& Ballegooijen, A. A.~v. 2005, The Astrophysical Journal
  Supplement Series, 156, 265, \dodoi{10.1086/426507}

\bibitem[{Cuesta {et~al.}(2022)Cuesta, Chhiber, Roy, Goodwill, Pecora, Jarosik,
  Matthaeus, Parashar, \& Bandyopadhyay}]{cuesta_isotropization_2022}
Cuesta, M.~E., Chhiber, R., Roy, S., {et~al.} 2022, The Astrophysical Journal
  Letters, 932, L11, \dodoi{10.3847/2041-8213/ac73fd}

\bibitem[{Denskat \& Neubauer(1982)}]{denskat_statistical_1982}
Denskat, K.~U., \& Neubauer, F.~M. 1982, Journal of Geophysical Research: Space
  Physics, 87, 2215, \dodoi{10.1029/JA087iA04p02215}

\bibitem[{Dmitruk \& Matthaeus(2007)}]{dmitruk_low-frequency_2007}
Dmitruk, P., \& Matthaeus, W.~H. 2007, Physical Review E, 76, 036305,
  \dodoi{10.1103/PhysRevE.76.036305}

\bibitem[{Dudok~de Wit {et~al.}(2020)Dudok~de Wit, Krasnoselskikh, Bale,
  Bonnell, Bowen, Chen, Froment, Goetz, Harvey, Jagarlamudi, Larosa, MacDowall,
  Malaspina, Matthaeus, Pulupa, Velli, \&
  Whittlesey}]{dudok_de_wit_switchbacks_2020}
Dudok~de Wit, T., Krasnoselskikh, V.~V., Bale, S.~D., {et~al.} 2020, The
  Astrophysical Journal Supplement Series, 246, 39,
  \dodoi{10.3847/1538-4365/ab5853}

\bibitem[{Farrell {et~al.}(2020)Farrell, MacDowall, Gruesbeck, Bale, \&
  Kasper}]{farrell_magnetic_2020}
Farrell, W.~M., MacDowall, R.~J., Gruesbeck, J.~R., Bale, S.~D., \& Kasper,
  J.~C. 2020, The Astrophysical Journal Supplement Series, 249, 28,
  \dodoi{10.3847/1538-4365/ab9eba}

\bibitem[{Forman {et~al.}(2011)Forman, Wicks, \&
  Horbury}]{forman_detailed_2011}
Forman, M.~A., Wicks, R.~T., \& Horbury, T.~S. 2011, The Astrophysical Journal,
  733, 76, \dodoi{10.1088/0004-637X/733/2/76}

\bibitem[{Fox {et~al.}(2016)Fox, Velli, Bale, Decker, Driesman, Howard, Kasper,
  Kinnison, Kusterer, Lario, Lockwood, McComas, Raouafi, \&
  Szabo}]{fox_solar_2016}
Fox, N.~J., Velli, M.~C., Bale, S.~D., {et~al.} 2016, Space Science Reviews,
  204, 7, \dodoi{10.1007/s11214-015-0211-6}

\bibitem[{Fu {et~al.}(2018)Fu, Li, Guo, Li, \&
  Roytershteyn}]{fu_parametric_2018}
Fu, X., Li, H., Guo, F., Li, X., \& Roytershteyn, V. 2018, The Astrophysical
  Journal, 855, 139, \dodoi{10.3847/1538-4357/aaacd6}

\bibitem[{Goldreich \& Sridhar(1995)}]{goldreich_toward_1995}
Goldreich, P., \& Sridhar, S. 1995, The Astrophysical Journal, 438, 763,
  \dodoi{10.1086/175121}

\bibitem[{Goldreich \& Sridhar(1997)}]{goldreich_magnetohydrodynamic_1997}
---. 1997, {\textbackslash}apj, 485, 680, \dodoi{10.1086/304442}

\bibitem[{Harris {et~al.}(2020)Harris, Millman, van~der Walt, Gommers,
  Virtanen, Cournapeau, Wieser, Taylor, Berg, Smith, Kern, Picus, Hoyer, van
  Kerkwijk, Brett, Haldane, del R{\'i}o, Wiebe, Peterson, G{\'e}rard-Marchant,
  Sheppard, Reddy, Weckesser, Abbasi, Gohlke, \& Oliphant}]{harris_array_2020}
Harris, C.~R., Millman, K.~J., van~der Walt, S.~J., {et~al.} 2020, Nature, 585,
  357, \dodoi{10.1038/s41586-020-2649-2}

\bibitem[{Heinemann \& Olbert(1980)}]{heinemann_non-wkb_1980}
Heinemann, M., \& Olbert, S. 1980, Journal of Geophysical Research: Space
  Physics, 85, 1311, \dodoi{10.1029/JA085iA03p01311}

\bibitem[{Hern{\'a}ndez {et~al.}(2021)Hern{\'a}ndez, Sorriso-Valvo,
  Bandyopadhyay, Chasapis, V{\'a}sconez, Marino, \&
  Pezzi}]{hernandez_impact_2021}
Hern{\'a}ndez, C.~S., Sorriso-Valvo, L., Bandyopadhyay, R., {et~al.} 2021, The
  Astrophysical Journal Letters, 922, L11, \dodoi{10.3847/2041-8213/ac36d1}

\bibitem[{Huang {et~al.}(2023)Huang, Kasper, Fisk, Larson, McManus, Chen,
  Martinovi{\'c}, Klein, Thomas, Liu, Maruca, Zhao, Chen, Hu, Jian, Verniero,
  Velli, Livi, Whittlesey, Rahmati, Romeo, Niembro, Paulson, Stevens, Case,
  Pulupa, Bale, \& Halekas}]{huang_structure_2023}
Huang, J., Kasper, J.~C., Fisk, L.~A., {et~al.} 2023, The {Structure} and
  {Origin} of {Switchbacks}: {Parker} {Solar} {Probe} {Observations},  arXiv,
  \dodoi{10.48550/arXiv.2301.10374}

\bibitem[{Huang {et~al.}(2022)Huang, Shi, Sioulas, \&
  Velli}]{huang_conservation_2022}
Huang, Z., Shi, C., Sioulas, N., \& Velli, M. 2022, The Astrophysical Journal,
  935, 60, \dodoi{10.3847/1538-4357/ac74c5}

\bibitem[{Hunter(2007)}]{hunter_matplotlib_2007}
Hunter, J.~D. 2007, Computing in Science \& Engineering, 9, 90,
  \dodoi{10.1109/MCSE.2007.55}

\bibitem[{Iroshnikov(1964)}]{iroshnikov_turbulence_1964}
Iroshnikov, P.~S. 1964, {\textbackslash}sovast, 7, 566

\bibitem[{Kasper {et~al.}(2016)Kasper, Abiad, Austin, Balat-Pichelin, Bale,
  Belcher, Berg, Bergner, Berthomier, Bookbinder, Brodu, Caldwell, Case,
  Chandran, Cheimets, Cirtain, Cranmer, Curtis, Daigneau, Dalton, Dasgupta,
  DeTomaso, Diaz-Aguado, Djordjevic, Donaskowski, Effinger, Florinski, Fox,
  Freeman, Gallagher, Gary, Gauron, Gates, Goldstein, Golub, Gordon, Gurnee,
  Guth, Halekas, Hatch, Heerikuisen, Ho, Hu, Johnson, Jordan, Korreck, Larson,
  Lazarus, Li, Livi, Ludlam, Maksimovic, McFadden, Marchant, Maruca, McComas,
  Messina, Mercer, Park, Peddie, Pogorelov, Reinhart, Richardson, Robinson,
  Rosen, Skoug, Slagle, Steinberg, Stevens, Szabo, Taylor, Tiu, Turin, Velli,
  Webb, Whittlesey, Wright, Wu, \& Zank}]{kasper_solar_2016}
Kasper, J.~C., Abiad, R., Austin, G., {et~al.} 2016, {\ss}r, 204, 131,
  \dodoi{10.1007/s11214-015-0206-3}

\bibitem[{Kasper {et~al.}(2021)Kasper, Klein, Lichko, Huang, Chen, Badman,
  Bonnell, Whittlesey, Livi, Larson, Pulupa, Rahmati, Stansby, Korreck,
  Stevens, Case, Bale, Maksimovic, Moncuquet, Goetz, Halekas, Malaspina,
  Raouafi, Szabo, MacDowall, Velli, Dudok~de Wit, \& Zank}]{kasper_parker_2021}
Kasper, J.~C., Klein, K.~G., Lichko, E., {et~al.} 2021, Physical Review
  Letters, 127, 255101, \dodoi{10.1103/PhysRevLett.127.255101}

\bibitem[{Keshner(1982)}]{keshner_1f_1982}
Keshner, M. 1982, Proceedings of the IEEE, 70, 212,
  \dodoi{10.1109/PROC.1982.12282}

\bibitem[{Kolmogorov(1941)}]{kolmogorov_local_1941}
Kolmogorov, A. 1941, Akademiia Nauk SSSR Doklady, 30, 301

\bibitem[{Kraichnan(1965)}]{kraichnan_inertialrange_1965}
Kraichnan, R. 1965, \dodoi{10.1063/1.1761412}

\bibitem[{Laker {et~al.}(2021)Laker, Horbury, Bale, Matteini, Woolley, Woodham,
  Badman, Pulupa, Kasper, Stevens, Case, \& Korreck}]{laker_statistical_2021}
Laker, R., Horbury, T.~S., Bale, S.~D., {et~al.} 2021, Astronomy and
  Astrophysics, 650, A1, \dodoi{10.1051/0004-6361/202039354}

\bibitem[{Lam {et~al.}(2015)Lam, Pitrou, \& Seibert}]{lam_numba_2015}
Lam, S.~K., Pitrou, A., \& Seibert, S. 2015, in Proceedings of the {Second}
  {Workshop} on the {LLVM} {Compiler} {Infrastructure} in {HPC} (Austin Texas:
  ACM), 1--6, \dodoi{10.1145/2833157.2833162}

\bibitem[{Larosa {et~al.}(2021)Larosa, Krasnoselskikh, Wit, Agapitov, Froment,
  Jagarlamudi, Velli, Bale, Case, Goetz, Harvey, Kasper, Korreck, Larson,
  MacDowall, Malaspina, Pulupa, Revillet, \& Stevens}]{larosa_switchbacks_2021}
Larosa, A., Krasnoselskikh, V., Wit, T. D.~d., {et~al.} 2021, Astronomy \&
  Astrophysics, 650, A3, \dodoi{10.1051/0004-6361/202039442}

\bibitem[{Livi {et~al.}(2022)Livi, Larson, Kasper, Abiad, Case, Klein, Curtis,
  Dalton, Stevens, Korreck, Ho, Robinson, Tiu, Whittlesey, Verniero, Halekas,
  McFadden, Marckwordt, Slagle, Abatcha, Rahmati, \& McManus}]{livi_solar_2022}
Livi, R., Larson, D.~E., Kasper, J.~C., {et~al.} 2022, The Astrophysical
  Journal, 938, 138, \dodoi{10.3847/1538-4357/ac93f5}

\bibitem[{Magyar \& Doorsselaere(2022)}]{magyar_phase_2022}
Magyar, N., \& Doorsselaere, T.~V. 2022, The Astrophysical Journal, 938, 98,
  \dodoi{10.3847/1538-4357/ac8b81}

\bibitem[{Mallet \& Chandran(2021)}]{mallet_exact_2021}
Mallet, A., \& Chandran, B. D.~G. 2021, Journal of Plasma Physics, 87,
  175870601, \dodoi{10.1017/S0022377821000970}

\bibitem[{Mallet {et~al.}(2021)Mallet, Squire, Chandran, Bowen, \&
  Bale}]{mallet_evolution_2021}
Mallet, A., Squire, J., Chandran, B. D.~G., Bowen, T., \& Bale, S.~D. 2021, The
  Astrophysical Journal, 918, 62, \dodoi{10.3847/1538-4357/ac0c12}

\bibitem[{Martinovi{\'c} {et~al.}(2021)Martinovi{\'c}, Klein, Huang, Chandran,
  Kasper, Lichko, Bowen, Chen, Matteini, Stevens, Case, \&
  Bale}]{martinovic_multiscale_2021}
Martinovi{\'c}, M.~M., Klein, K.~G., Huang, J., {et~al.} 2021, The
  Astrophysical Journal, 912, 28, \dodoi{10.3847/1538-4357/abebe5}

\bibitem[{Matteini {et~al.}(2010)Matteini, Landi, Del~Zanna, Velli, \&
  Hellinger}]{matteini_parametric_2010}
Matteini, L., Landi, S., Del~Zanna, L., Velli, M., \& Hellinger, P. 2010,
  Geophysical Research Letters, 37, L20101, \dodoi{10.1029/2010GL044806}

\bibitem[{Matteini {et~al.}(2018)Matteini, Stansby, Horbury, \&
  Chen}]{matteini_1_2018}
Matteini, L., Stansby, D., Horbury, T.~S., \& Chen, C. H.~K. 2018, The
  Astrophysical Journal, 869, L32, \dodoi{10.3847/2041-8213/aaf573}

\bibitem[{Matthaeus {et~al.}(2007)Matthaeus, Breech, Dmitruk, Bemporad,
  Poletto, Velli, \& Romoli}]{matthaeus_density_2007}
Matthaeus, W.~H., Breech, B., Dmitruk, P., {et~al.} 2007, The Astrophysical
  Journal, 657, L121, \dodoi{10.1086/513075}

\bibitem[{Matthaeus \& Goldstein(1986)}]{matthaeus_low-frequency_1986}
Matthaeus, W.~H., \& Goldstein, M.~L. 1986, Physical Review Letters, 57, 495,
  \dodoi{10.1103/PhysRevLett.57.495}

\bibitem[{Matthaeus {et~al.}(1982)Matthaeus, Goldstein, \&
  Smith}]{matthaeus_evaluation_1982}
Matthaeus, W.~H., Goldstein, M.~L., \& Smith, C. 1982, Physical Review Letters,
  48, 1256, \dodoi{10.1103/PhysRevLett.48.1256}

\bibitem[{Meng {et~al.}(2022)Meng, Liu, Chen, \& Wang}]{meng_analysis_2022}
Meng, M.-M., Liu, Y.~D., Chen, C., \& Wang, R. 2022, Research in Astronomy and
  Astrophysics, 22, 035018, \dodoi{10.1088/1674-4527/ac49e4}

\bibitem[{Moncuquet {et~al.}(2020)Moncuquet, Meyer-Vernet, Issautier, Pulupa,
  Bonnell, Bale, Wit, Goetz, Griton, Harvey, MacDowall, Maksimovic, \&
  Malaspina}]{moncuquet_first_2020}
Moncuquet, M., Meyer-Vernet, N., Issautier, K., {et~al.} 2020, The
  Astrophysical Journal Supplement Series, 246, 44,
  \dodoi{10.3847/1538-4365/ab5a84}

\bibitem[{Monin \& Jaglom(1987)}]{monin_statistical_1987}
Monin, A.~S., \& Jaglom, A.~M. 1987, Statistical fluid mechanics. 2, 3rd edn.
  (Cambridge, Mass.: MIT Pr)

\bibitem[{Montroll \& Shlesinger(1982)}]{montroll_1f_1982}
Montroll, E.~W., \& Shlesinger, M.~F. 1982, Proceedings of the National Academy
  of Sciences of the United States of America, 79, 3380,
  \dodoi{10.1073/pnas.79.10.3380}

\bibitem[{Mozer {et~al.}(2020)Mozer, Agapitov, Bale, Bonnell, Goetz, Goodrich,
  Gore, Harvey, Kellogg, Malaspina, Pulupa, \& Schumm}]{mozer_time_2020}
Mozer, F.~S., Agapitov, O.~V., Bale, S.~D., {et~al.} 2020, The Astrophysical
  Journal Supplement Series, 246, 50, \dodoi{10.3847/1538-4365/ab5e4b}

\bibitem[{Nakagawa \& Levine(1974)}]{nakagawa_dynamics_1974}
Nakagawa, Y., \& Levine, R.~H. 1974, The Astrophysical Journal, 190, 441,
  \dodoi{10.1086/152896}

\bibitem[{Parker(1958)}]{parker_dynamics_1958}
Parker, E.~N. 1958, The Astrophysical Journal, 128, 664, \dodoi{10.1086/146579}

\bibitem[{Perez {et~al.}(2021{\natexlab{a}})Perez, Bourouaine, Chen, \&
  Raouafi}]{perez_applicability_2021}
Perez, J.~C., Bourouaine, S., Chen, C. H.~K., \& Raouafi, N.~E.
  2021{\natexlab{a}}, Astronomy \& Astrophysics, 650, A22,
  \dodoi{10.1051/0004-6361/202039879}

\bibitem[{Perez {et~al.}(2021{\natexlab{b}})Perez, Chandran, Klein, \&
  Martinovi{\'c}}]{perez_how_2021}
Perez, J.~C., Chandran, B. D.~G., Klein, K.~G., \& Martinovi{\'c}, M.~M.
  2021{\natexlab{b}}, Journal of Plasma Physics, 87, 905870218,
  \dodoi{10.1017/S0022377821000167}

\bibitem[{Podesta(2009)}]{podesta_dependence_2009}
Podesta, J.~J. 2009, The Astrophysical Journal, 698, 986,
  \dodoi{10.1088/0004-637X/698/2/986}

\bibitem[{Pulupa {et~al.}(2017)Pulupa, Bale, Bonnell, Bowen, Carruth, Goetz,
  Gordon, Harvey, Maksimovic, Mart{\'i}nez-Oliveros, Moncuquet, Saint-Hilaire,
  Seitz, \& Sundkvist}]{pulupa_solar_2017}
Pulupa, M., Bale, S.~D., Bonnell, J.~W., {et~al.} 2017, Journal of Geophysical
  Research: Space Physics, 122, 2836, \dodoi{10.1002/2016JA023345}

\bibitem[{Raouafi {et~al.}(2023)Raouafi, Matteini, Squire, Badman, Velli,
  Klein, Chen, Matthaeus, Szabo, Linton, Allen, Szalay, Bruno, Decker,
  Akhavan-Tafti, Agapitov, Bale, Bandyopadhyay, Battams, Ber{\v c}i{\v c},
  Bourouaine, Bowen, Cattell, Chandran, Chhiber, Cohen,
  D{\textquoteright}Amicis, Giacalone, Hess, Howard, Horbury, Jagarlamudi,
  Joyce, Kasper, Kinnison, Laker, Liewer, Malaspina, Mann, McComas,
  Niembro-Hernandez, Nieves-Chinchilla, Panasenco, Pokorn{\'y}, Pusack, Pulupa,
  Perez, Riley, Rouillard, Shi, Stenborg, Tenerani, Verniero, Viall, Vourlidas,
  Wood, Woodham, \& Woolley}]{raouafi_parker_2023}
Raouafi, N.~E., Matteini, L., Squire, J., {et~al.} 2023, Space Science Reviews,
  219, 8, \dodoi{10.1007/s11214-023-00952-4}

\bibitem[{R{\'e}ville {et~al.}(2018)R{\'e}ville, Tenerani, \&
  Velli}]{reville_parametric_2018}
R{\'e}ville, V., Tenerani, A., \& Velli, M. 2018, The Astrophysical Journal,
  866, 38, \dodoi{10.3847/1538-4357/aadb8f}

\bibitem[{Shi {et~al.}(2022{\natexlab{a}})Shi, Velli, Bale, R{\'e}ville,
  Maksimovi{\'c}, \& Dakeyo}]{shi_acceleration_2022}
Shi, C.~., Velli, M., Bale, S.~D., {et~al.} 2022{\natexlab{a}}, Physics of
  Plasmas, 29, 122901, \dodoi{10.1063/5.0124703}

\bibitem[{Shi {et~al.}(2021)Shi, Velli, Panasenco, Tenerani, R{\'e}ville, Bale,
  Kasper, Korreck, Bonnell, de~Wit, Malaspina, Goetz, Harvey, MacDowall,
  Pulupa, Case, Larson, Verniero, Livi, Stevens, Whittlesey, Maksimovic, \&
  Moncuquet}]{shi_alfvenic_2021}
Shi, C., Velli, M., Panasenco, O., {et~al.} 2021, Astronomy \& Astrophysics,
  650, A21, \dodoi{10.1051/0004-6361/202039818}

\bibitem[{Shi {et~al.}(2022{\natexlab{b}})Shi, Panasenco, Velli, Tenerani,
  Verniero, Sioulas, Huang, Brosius, Bale, \& Klein}]{shi_patches_2022}
Shi, C., Panasenco, O., Velli, M., {et~al.} 2022{\natexlab{b}}, The
  Astrophysical Journal, 934, 152

\bibitem[{Shi {et~al.}(2017)Shi, Li, Xiao, \& Wang}]{shi_parametric_2017}
Shi, M., Li, H., Xiao, C., \& Wang, X. 2017, The Astrophysical Journal, 842,
  63, \dodoi{10.3847/1538-4357/aa71b6}

\bibitem[{Sioulas {et~al.}(2022)Sioulas, Huang, Velli, Chhiber, Cuesta, Shi,
  Matthaeus, Bandyopadhyay, Vlahos, \& Bowen}]{sioulas_magnetic_2022}
Sioulas, N., Huang, Z., Velli, M., {et~al.} 2022, arXiv preprint
  arXiv:2206.00871

\bibitem[{Sioulas {et~al.}(2023{\natexlab{a}})Sioulas, Huang, Shi, Velli,
  Tenerani, Vlahos, Bowen, Bale, Bonnell, Harvey, Larson, Pulupa, Livi,
  Woodham, Horbury, Stevens, de~Wit, MacDowall, Malaspina, Goetz, Huang,
  Kasper, Owen, Maksimovi{\'c}, Louarn, \& Fedorov}]{sioulas_magnetic_2023}
Sioulas, N., Huang, Z., Shi, C., {et~al.} 2023{\natexlab{a}}, The Astrophysical
  Journal Letters, 943, L8, \dodoi{10.3847/2041-8213/acaeff}

\bibitem[{Sioulas {et~al.}(2023{\natexlab{b}})Sioulas, Velli, Huang, Shi,
  Bowen, Chandran, Liodis, Bale, Horbury, de~Wit, Larson, Kasper, Owen,
  Stevens, Case, Pulupa, Bonnell, Goetz, Harvey, \&
  MacDowall}]{sioulas_evolution_2023}
Sioulas, N., Velli, M., Huang, Z., {et~al.} 2023{\natexlab{b}}, On the
  evolution of the {Anisotropic} {Scaling} of {Magnetohydrodynamic}
  {Turbulence} in the {Inner} {Heliosphere},  arXiv,
  \dodoi{10.48550/arXiv.2301.03896}

\bibitem[{Sridhar \& Goldreich(1994)}]{sridhar_toward_1994}
Sridhar, S., \& Goldreich, P. 1994, The Astrophysical Journal, 432, 612,
  \dodoi{10.1086/174600}

\bibitem[{Taylor(1938)}]{taylor_spectrum_1938}
Taylor, G.~I. 1938, Proceedings of the Royal Society of London Series A, 164,
  476, \dodoi{10.1098/rspa.1938.0032}

\bibitem[{Taylor(1986)}]{taylor_relaxation_1986}
Taylor, J.~B. 1986, Rev. Mod. Phys., 58, 741, \dodoi{10.1103/RevModPhys.58.741}

\bibitem[{Team(2023)}]{team_pandas-devpandas_2023}
Team, T. P.~D. 2023, pandas-dev/pandas: {Pandas},  Zenodo,
  \dodoi{10.5281/ZENODO.3509134}

\bibitem[{Telloni {et~al.}(2021)Telloni, Sorriso-Valvo, Woodham, Panasenco,
  Velli, Carbone, Zank, Bruno, Perrone, Nakanotani, Shi, D'Amicis, De~Marco,
  Jagarlamudi, Steinvall, Marino, Adhikari, Zhao, Liang, Tenerani, Laker,
  Horbury, Bale, Pulupa, Malaspina, MacDowall, Goetz, de~Wit, Harvey, Kasper,
  Korreck, Larson, Case, Stevens, Whittlesey, Livi, Owen, Livi, Louarn,
  Antonucci, Romoli, O'Brien, Evans, \& Angelini}]{telloni_evolution_2021}
Telloni, D., Sorriso-Valvo, L., Woodham, L.~D., {et~al.} 2021,
  {\textbackslash}apjl, 912, L21, \dodoi{10.3847/2041-8213/abf7d1}

\bibitem[{Tenerani {et~al.}(2021)Tenerani, Sioulas, Matteini, Panasenco, Shi,
  \& Velli}]{tenerani_evolution_2021}
Tenerani, A., Sioulas, N., Matteini, L., {et~al.} 2021, The Astrophysical
  Journal Letters, 919, L31, \dodoi{10.3847/2041-8213/ac2606}

\bibitem[{Tenerani \& Velli(2013)}]{tenerani_parametric_2013}
Tenerani, A., \& Velli, M. 2013, Journal of Geophysical Research: Space
  Physics, 118, 7507, \dodoi{10.1002/2013JA019293}

\bibitem[{Tenerani {et~al.}(2017)Tenerani, Velli, \&
  Hellinger}]{tenerani_parametric_2017}
Tenerani, A., Velli, M., \& Hellinger, P. 2017, The Astrophysical Journal, 851,
  99, \dodoi{10.3847/1538-4357/aa9bef}

\bibitem[{Tenerani {et~al.}(2020)Tenerani, Velli, Matteini, Reville, Shi, Bale,
  Kasper, Bonnell, Case, de~Wit, Goetz, Harvey, Klein, Korreck, Larson, Livi,
  MacDowall, Malaspina, Pulupa, Stevens, \&
  Whittlesey}]{tenerani_magnetic_2020}
Tenerani, A., Velli, M., Matteini, L., {et~al.} 2020, The Astrophysical
  Journal, 9

\bibitem[{Tu \& Marsch(1995)}]{tu_mhd_1995}
Tu, C.~Y., \& Marsch, E. 1995, Space Science Reviews, 73, 1,
  \dodoi{10.1007/BF00748891}

\bibitem[{Velli {et~al.}(1989)Velli, Grappin, \&
  Mangeney}]{velli_turbulent_1989}
Velli, M., Grappin, R., \& Mangeney, A. 1989, Physical Review Letters, 63,
  1807, \dodoi{10.1103/PhysRevLett.63.1807}

\bibitem[{Velli {et~al.}(1991)Velli, Grappin, \& Mangeney}]{velli_waves_1991}
---. 1991, Geophysical \& Astrophysical Fluid Dynamics, 62, 101,
  \dodoi{10.1080/03091929108229128}

\bibitem[{Verdini {et~al.}(2012)Verdini, Grappin, Pinto, \&
  Velli}]{verdini_origin_2012}
Verdini, A., Grappin, R., Pinto, R., \& Velli, M. 2012, The Astrophysical
  Journal, 750, L33, \dodoi{10.1088/2041-8205/750/2/L33}

\bibitem[{Verdini \& Velli(2007)}]{verdini_alfven_2007}
Verdini, A., \& Velli, M. 2007, The Astrophysical Journal, 662, 669,
  \dodoi{10.1086/510710}

\bibitem[{Virtanen {et~al.}(2020)Virtanen, Gommers, Oliphant, Haberland, Reddy,
  Cournapeau, Burovski, Peterson, Weckesser, Bright, van~der Walt, Brett,
  Wilson, Millman, Mayorov, Nelson, Jones, Kern, Larson, Carey, Polat, Feng,
  Moore, VanderPlas, Laxalde, Perktold, Cimrman, Henriksen, Quintero, Harris,
  Archibald, Ribeiro, Pedregosa, \& van Mulbregt}]{virtanen_scipy_2020}
Virtanen, P., Gommers, R., Oliphant, T.~E., {et~al.} 2020, Nature Methods, 17,
  261, \dodoi{10.1038/s41592-019-0686-2}

\bibitem[{Whang(1973)}]{whang_alfven_1973}
Whang, Y.~C. 1973, Journal of Geophysical Research, 78, 7221,
  \dodoi{10.1029/JA078i031p07221}

\bibitem[{Woolley {et~al.}(2020)Woolley, Matteini, Horbury, Bale, Woodham,
  Laker, Alterman, Bonnell, Case, Kasper, Klein, Martinovi{\'c}, \&
  Stevens}]{woolley_proton_2020}
Woolley, T., Matteini, L., Horbury, T.~S., {et~al.} 2020, Monthly Notices of
  the Royal Astronomical Society, 498, 5524, \dodoi{10.1093/mnras/staa2770}

\bibitem[{Zank {et~al.}(2022)Zank, Zhao, Adhikari, Telloni, Kasper, Stevens,
  Rahmati, \& Bale}]{zank_turbulence_2022}
Zank, G.~P., Zhao, L.-L., Adhikari, L., {et~al.} 2022, The Astrophysical
  Journal Letters, 926, L16, \dodoi{10.3847/2041-8213/ac51da}

\bibitem[{Zanna {et~al.}(2015)Zanna, Matteini, Landi, Verdini, \&
  Velli}]{zanna_parametric_2015}
Zanna, L.~D., Matteini, L., Landi, S., Verdini, A., \& Velli, M. 2015, Journal
  of Plasma Physics, 81, 325810102, \dodoi{10.1017/S0022377814000579}

\end{thebibliography}
\bibliographystyle{aasjournal}

\end{CJK*}
\end{document}